\documentclass{article}
\usepackage{amsmath}
\usepackage{amssymb}

\newtheorem{theorem}{Theorem}

\newtheorem{corollary}[theorem]{Corollary}

\newenvironment{proof}[1][Proof]{\noindent\textbf{#1.} }{\ \rule{0.5em}{0.5em}}
\input{tcilatex}
\begin{document}

\title{On $f(R)$ theories equivalent to general relativity for a specific
form of the Ricci scalar}
\author{Amine Benachour \\
Laboratoire de Physique Th\'{e}orique, \ \\
Facult\'{e} des Sciences Exactes, \\
Universit\'{e} Fr\`{e}res Mentouri Constantine 1, \\
Route de A\"{\i}n El-Bey,\\
Constantine 25017, Algeria.\\
a.benachour@umc.edu.dz}
\maketitle

\begin{abstract}
We prove the necessary and the sufficient condition on the $f(R)$ function
leading to the equivalence of general relativity and the $f(R)$ theory in
the sense that for a spacetime with specific Ricci scalar, GR and the $f(R)$
theory will necessarily have the same solutions. We show how this condition
and its consequences allow to get such functions and give the possibility to
fixe conditions on the parameters of any class of $f(R)$ functions so that
GR and this class of theories will have exactly the same solutions. The
well-known model $R+\alpha \beta ^{1-p}R^{p}$ for dark energy is presented
as an illustration, as well as the FLRW metric where it's shown that there
exists only one class of $f(R)$ theories which has exactly the same GR scale
factor.
\end{abstract}



\section{Introduction}

As most of the physical laws, General relativity considered by many
physicists as the best theory for the description of gravitation can be
derived from the fascinating variational principle. Indeed Einstein's field
equations%
\begin{equation}
R_{\mu \nu }-\frac{1}{2}Rg_{\mu \nu }+\Lambda g_{\mu \nu }=\kappa T_{\mu \nu
};\kappa =\frac{8\pi G}{c^{4}},  \label{1}
\end{equation}%
are easily obtained from the variation of the Hilbert-Einstein action 
\begin{equation}
S=\frac{1}{2\kappa }\int \sqrt{-g}\left( R-2\Lambda \right) d^{4}x+S_{m}.
\label{2}
\end{equation}%
The idea of what's called $f(R)$ modified theory of gravity is based on the
generalization of the Hilbert-Einstein action by writing $S$ in more general
form 
\begin{equation}
S=\frac{1}{2\kappa }\int \sqrt{-g}f\left( R\right) d^{4}x+S_{m},  \label{3}
\end{equation}%
enabling to deduce the $f(R)$ field equations%
\begin{equation}
f^{^{\prime }}\left( R\right) R_{\mu \nu }-\frac{1}{2}f\left( R\right)
g_{\mu \nu }-\nabla _{\mu }\nabla _{\nu }f^{^{\prime }}\left( R\right)
+\square f^{\prime }\left( R\right) g_{\mu \nu }=\kappa T_{\mu \nu },
\label{4}
\end{equation}%
after a variation with respect to the metric $g_{\mu \nu },$ where $%
f^{^{\prime }}\left( R\right) =\frac{df\left( R\right) }{dR}$ and $\nabla
_{\mu }$represents the covariant derivative. Thereby, for any given $f(R)$
function we get an $f(R)$ theory and general relativity becomes the $f(R)$
theory corresponding to the particular case for which $f(R)=R-2\Lambda $,
among an infinity of possible choices for every $f(R)$ function. Dealing
with any modified theory of gravity is a way of going beyond Einstein theory%
\cite{far} intended primarily for solving many current problems in cosmology
like the question of dark energy, where $f(R)$ theory by the mean of its
different proposed models\cite{men} is an alternative and a candidate to
explain an expanding accelerating universe without dark energy. However, the
predictions of any proposed $f(R)$ theory must be in agreement with
observations in order to be acceptable as model for dark energy, which
translates to mathematical conditions on the $f(R)$ functions restricting
the choice of the functions $f(R)$ to a family of theories. Independently
from whether the $f(R)$ function satisfies the required conditions or not,
it's rightful to ask the question: Is it possible that GR and an $f(R)\neq $ 
$R-2\Lambda $ theory have the same solutions when the two theories are
concerned with the same physical context?. If so, when?. By solutions we
mean the metric tensor $g_{\mu \nu }$ and "the same physical context" means
the same stress-energy tensor $T_{\mu \nu }$ describing a specific type of
matter for both field equations (\ref{1}), (\ref{4}). Actually there is two
motivations for asking this question: The first one is the difficulty of
solving exactly $f(R)$ field equations; so before envisaging any technique
of resolution, it's worth to know if the GR solution still to be an $f(R)$
solution since GR field equations are easier to solve than $f(R)$ ones, and
most of the interesting GR exact solutions are well- known since long time
ago\cite{stef}. The second motivation is to measure how much the proposed
modified theory is so different from general relativity; it has no sense to
build more complicated theories seeming diffrent, in the end it turns out
that the relevant solutions of these theories are those of general
relativity; so any investigation for understanding the relation between $%
f(R) $ and GR solutions is highly suitable. The aim of this work is to give
an answer to the previous question by determining first all the mathematical
conditions on the function $f(R)$ for which GR and $f(R)$ theory will have
the same solutions, then we will show how one may exploit these conditions
by applying them on concrete examples.

\section{The conditions required on $f(R)$ functions to have common
solutions GR - $f(R)$ theory}

As we have pointed out, our objective is to answer the question: when will
GR and $f(R)$ theory have the same solutions? We will first start to answer
this question in the general case when there is no assumptions on the Ricci
scalar, then we will consider the particular case of spacetimes with
constant Ricci scalar, not just for its simplicity but also for its physical
importance.

\subsection{The general case (R is arbitrary)}

In order to give an answer to the previous question, the key idea consists
in bring back Einstein's equations inside $f(R)$ field equations(\ref{4})
which can be written in the equivalent form%
\begin{eqnarray}
&&f^{\prime }(R)\left[ R_{\mu \nu }-\frac{R}{2}g_{\mu \nu }+\Lambda g_{\mu
\nu }-\kappa T_{\mu \nu }\right]  \label{40} \\
&&+f^{\prime }(R)\left[ \frac{R}{2}g_{\mu \nu }-\Lambda g_{\mu \nu }+\kappa
T_{\mu \nu }\right]  \notag \\
&&-\frac{1}{2}f\left( R\right) g_{\mu \nu }-\nabla _{\mu }\nabla _{\nu
}f^{^{\prime }}\left( R\right)  \notag \\
+\square f^{\prime }\left( R\right) g_{\mu \nu } &=&\kappa T_{\mu \nu }, 
\notag
\end{eqnarray}%
with $f^{\prime }\left( R\right) =\left[ \frac{df\left( R\right) }{dR}\right]
,$ then we can state the following theorem:

\begin{theorem}
\emph{A metric }$g_{\mu \nu }$\emph{\ is solution of GR and }$f(R)$\emph{\
field equations if and only if it's at least solution of one of the two
theories and } 
\begin{equation}
\left[ f^{\prime }\left( R\right) -1\right] R_{\mu \nu }-\frac{1}{2}\left[
f(R)-\left( R-2\Lambda \right) -2\square f^{\prime }\left( R\right) \right]
g_{\mu \nu }-\nabla _{\mu }\nabla _{\nu }f^{^{\prime }}\left( R\right) =0.
\label{11}
\end{equation}
\end{theorem}

\begin{proof}
Let $g_{\mu \nu }$ be a solution of GR and $f(R)$ theory; so $g_{\mu \nu }$
satisfies $f(R)$ field equation and GR ones, thus (\ref{40}) is satisfied
and its first term vanishes so that 
\begin{equation}
\left[ \left( \frac{R}{2}-\Lambda \right) f^{^{\prime }}\left( R\right) -%
\frac{1}{2}f(R)+\square f^{\prime }\left( R\right) \right] g_{\mu \nu }+%
\left[ f^{\prime }\left( R\right) -1\right] \kappa T_{\mu \nu }-\nabla _{\mu
}\nabla _{\nu }f^{^{\prime }}\left( R\right) =0.  \label{001}
\end{equation}%
Since $\kappa T_{\mu \nu }=R_{\mu \nu }-\frac{1}{2}Rg_{\mu \nu }+\Lambda
g_{\mu \nu },$ (\ref{001}) can be written as $\left[ f^{\prime }\left(
R\right) -1\right] R_{\mu \nu }-\frac{1}{2}\left[ f(R)-\left( R-2\Lambda
\right) -2\square f^{\prime }\left( R\right) \right] g_{\mu \nu }-\nabla
_{\mu }\nabla _{\nu }f^{^{\prime }}\left( R\right) =0$ which is the
necessary condition (\ref{11}); in other words if $g_{\mu \nu }$ is a
solution of both theories, we will necessarily have (\ref{11}). Let's prove
that the converse is also true by supposing a metric $g_{\mu \nu }$ which is
at least solution of one of the two theories satisfying at the same time the
condition (\ref{11}), to show that this is a sufficient condition for the
metric to be a solution for both theories. We will first start with the case
where $g_{\mu \nu }$ is a solution of GR. By adding GR equations and the
condition (\ref{11}), we get $\left[ R_{\mu \nu }-\frac{R}{2}g_{\mu \nu
}+\Lambda g_{\mu \nu }-\kappa T_{\mu \nu }\right] +\left[ f^{\prime }\left(
R\right) -1\right] R_{\mu \nu }-\frac{1}{2}\left[ f(R)-\left( R-2\Lambda
\right) -2\square f^{\prime }\left( R\right) \right] g_{\mu \nu }-\nabla
_{\mu }\nabla _{\nu }f^{^{\prime }}\left( R\right) =0,$ after simplification
we obtain $f(R)$ field equation (\ref{4}) which show that $g_{\mu \nu }$ is
also solution of the $f(R)$ theory. Now if we just know that $g_{\mu \nu }$
is a solution of $f(R)$ theory, satisfying (\ref{4}) and (\ref{11}), we can
with the same manner show that this is also sufficient to prove that this
metric will be a GR solution. Indeed; since $g_{\mu \nu }$ is solution of $%
f(R)$ field equations (\ref{4}) it will obviously be solution of equations (%
\ref{40}) because these ones are just an equivalent manner to write $f(R)$
equations by adding and subtracting the same quantity, namely Einstein
equations, on the other hand the term $f^{\prime }\left( R\right) \left[ 
\frac{R}{2}g_{\mu \nu }-\Lambda g_{\mu \nu }+\kappa T_{\mu \nu }\right] -%
\frac{1}{2}f\left( R\right) g_{\mu \nu }-\nabla _{\mu }\nabla _{\nu
}f^{\prime }\left( R\right) +\square f^{\prime }\left( R\right) g_{\mu \nu
}-\kappa T_{\mu \nu }$ vanishes because it's none other than the condition (%
\ref{11}) when $\kappa T_{\mu \nu }$ is replaced by $R_{\mu \nu }-\frac{1}{2}%
Rg_{\mu \nu }+\Lambda g_{\mu \nu }$; in fact we are allowed to make such
substitution despite the fact that we didn't yet proved that $g_{\mu \nu }$
is a GR solution because we are considering the same stress-energy tensor $%
T_{\mu \nu },$ for the two theories. Therfore $f^{^{\prime }}\left( R\right) %
\left[ R_{\mu \nu }-\frac{R}{2}g_{\mu \nu }+\Lambda g_{\mu \nu }-\kappa
T_{\mu \nu }\right] =0.$ As the $f(R)$ function isn't a constant but in
general it has a specific expresion in term of $R$ which in turn depends on
the system of coordinates; the derivative $f^{^{\prime }}\left( R\right) $
cannot vanish identically, therefore $R_{\mu \nu }-\frac{R}{2}g_{\mu \nu
}+\Lambda g_{\mu \nu }-\kappa T_{\mu \nu }=0,$ which prove that $g_{\mu \nu }
$ is a GR solution. Yet there exists the possibility to have a zero
derivative in the particular case of spacetimes with constant Ricci scalar
when this one is completely independent from the system of coordinates i.e.
when $R$ has the same constant value at every point of the spacetime like
with De Sitter spacetime. For instance if $f(R)=R^{2}-8\Lambda ,$ and the
metric leads to the constant Ricci $R=g_{\mu \nu }R^{\mu \nu }=4\Lambda ,$
then $f^{^{\prime }}\left( R\right) =f^{^{\prime }}\left( 4\Lambda \right) =%
\left[ 2R-8\Lambda \right] _{R=4\Lambda }=0.$ In fact this is does not
constitute any problem because the condition is strong enough to be valid
even when $f^{^{\prime }}\left( R\right) =0.$ Indeed in this case, $f(R)$
field equations are given by 
\begin{equation}
\frac{1}{2}f(R)g_{\mu \nu }+\kappa T_{\mu \nu }=0,  \label{004}
\end{equation}%
which could be written in the equivalent form 
\begin{equation}
\left[ R_{\mu \nu }-\frac{R}{2}g_{\mu \nu }+\Lambda g_{\mu \nu }-\kappa
T_{\mu \nu }\right] +\left[ R_{\mu \nu }+\frac{1}{2}\left( f(R)-\left(
R-2\Lambda \right) \right) g_{\mu \nu }\right] =0,  \label{005}
\end{equation}%
after adding and subtracting Einstein equations. We clearly see that the
first term between brackets is GR equations and the second one is the
condition (\ref{11}) in the particular case where $f^{^{\prime }}\left(
R\right) =0,$ so if $g_{\mu \nu }$ is an $f(R)$ solution and the condition (%
\ref{11}) is satisfied, the second term vanishes, then $g_{\mu \nu }$will
necessarily be a GR solution.
\end{proof}

We thus conclude that for all the $f(R)$ theories and the Ricci scalar $R$,
the condition (\ref{11}) constitutes a necessarily and a sufficient
condition for any metric which is solution of GR or the $f(R)$ theory, for
being simultaniousely a solution for the two theories.

\subsection{Specetimes with constant Ricci scalar ($R=R_{0}$)}

When one deal with a metric $g_{\mu \nu }$ describing a spacetime with
constant Ricci scalar, i.e. $g^{\mu \nu }R_{\mu \nu }=R_{0}$, where $R_{0}$
is a real number, for all the $f(R)$ functions, the derivative at $R_{0}$: $%
f^{\prime }\left( R_{0}\right) ,$ is always constant so the components of $%
\nabla _{\mu }\nabla _{\nu }f^{^{\prime }}\left( R_{0}\right) $ and the term 
$\square f^{\prime }\left( R_{0}\right) $ vanish and we can deduce from the
previous theorem, the following corollary,

\begin{corollary}
If $g_{\mu \nu }$ is a solution of GR or an $f(R)$ theory such as Ricci
scalar is a constant $R_{0}$; it's necessary and sufficient for this metric
to satisfy 
\begin{equation}
\left[ f^{\prime }\left( R_{0}\right) -1\right] R_{\mu \nu }-\frac{1}{2}%
\left[ f(R_{0})-\left( R_{0}-2\Lambda \right) \right] g_{\mu \nu }=0,
\label{006}
\end{equation}%
to be considered as a solution for both theories.
\end{corollary}

\begin{proof}
It's is just the particular case of (\ref{11}) when $R=R_{0}$ and\bigskip\ $%
\nabla _{\mu }\nabla _{\nu }f^{^{\prime }}\left( R_{0}\right) =0$, $\square
f^{\prime }\left( R_{0}\right) =0.$
\end{proof}

This is a condition which we can learn many things about the relation $f(R)$%
-GR, as we will show in what follows.

\begin{corollary}
For spacetimes with constant Ricci scalar $R_{0}$; General relativity and $%
f(R)$ theories such as, 
\begin{eqnarray}
f^{\prime }\left( R_{0}\right) &=&1  \label{12} \\
f\left( R_{0}\right) &=&R_{0}-2\Lambda ,  \notag
\end{eqnarray}%
are equivalent i.e. the GR and these theories have exactly the same
solutions for the spacetime of constant Ricci scalar $R_{0}$.
\end{corollary}

\begin{proof}
Let $g_{\mu \nu }$ be an $f(R)$ solution such as $g^{\mu \nu }R_{\mu \nu
}=R_{0}$. When Ricci scalar is constant, $f(R)$ field equations (\ref{4})
are given by 
\begin{equation}
f^{^{\prime }}\left( R_{0}\right) R_{\mu \nu }-\frac{1}{2}f\left(
R_{0}\right) g_{\mu \nu }=\kappa T_{\mu \nu },  \label{007}
\end{equation}%
so for $f(R)$ theories such as $f^{\prime }\left( R_{0}\right) =1,$ and $%
f\left( R_{0}\right) =R_{0}-2\Lambda ,$ the equations (\ref{007}) will be
identical to Einstein equations, which means that all the $f(R)$ solutions
such as $R=R_{0},$ are also the solutions of GR and vice versa. Another way
to prove this corollary is to see that if $f^{\prime }\left( R_{0}\right) =1$
and $f\left( R_{0}\right) =R_{0}-2\Lambda $ then the necessary and
sufficient condition (\ref{006}) for which a GR or an $f(R)$ metric is a
solution for the two theories; is satisfied.
\end{proof}

\begin{corollary}
The vaccum solution of GR is also a vacuum solution of an $f(R)$ theory if
and only if 
\begin{equation}
\Lambda f^{\prime }(4\Lambda )-\frac{1}{2}f(4\Lambda )=0.  \label{008}
\end{equation}
\end{corollary}

\begin{proof}
Let $g_{\mu \nu }$ be a vaccum solution of GR and $f(R)$ is a function such
as $\Lambda f^{\prime }(4\Lambda )-\frac{1}{2}f(4\Lambda )=0.$ As $g_{\mu
\nu }$ is a GR vaccum solution, $g^{\mu \nu }R_{\mu \nu }=4\Lambda ,$ \ so
Einstein field equations are given by $R_{\mu \nu }-\Lambda g_{\mu \nu }=0;$
thus $\left[ \Lambda f^{\prime }(4\Lambda )-\frac{1}{2}f(4\Lambda )\right]
g_{\mu \nu }-\left( R_{\mu \nu }-\Lambda g_{\mu \nu }\right) =0.$ If we
develop and we replace the term $f^{\prime }(4\Lambda )\Lambda g_{\mu \nu }$
by $f^{\prime }(4\Lambda )R_{\mu \nu },$we can rewrite this equation in the
equivalent form $\left[ f^{\prime }\left( 4\Lambda \right) -1\right] R_{\mu
\nu }-\frac{1}{2}\left[ f(4\Lambda )-2\Lambda \right] g_{\mu \nu }=0$, which
means that $g_{\mu \nu }$ is also a vaccum solution of $f(R)$ theory because
this equation is none other than the necessary and sufficient condition (\ref%
{006}) for which a GR solution will be an $f(R)$ one, in the particular case
where $R_{0}=4\Lambda $ i.e. the vaccum.
\end{proof}

\begin{corollary}
For Ricci flat manifolds, general relativity and $f(R)$ theories such as 
\begin{equation}
f(0)=-2\Lambda ,  \label{z}
\end{equation}
are equivalent, in other words all GR solutions describing manifolds with
zero Ricci tensor are solutions of $f(0)=-2\Lambda ,$ theories and vice
versa.
\end{corollary}

\begin{proof}
Let $g_{\mu \nu }$ be a metric such as $R_{\mu \nu }=0;$ so $R=0.$ According
to $f(R)$ field equations (\ref{007}); if $g_{\mu \nu }$ is a solution of an 
$f(R)$ theory such as $f(0)=-2\Lambda ,$ it satisfies: $\Lambda g_{\mu \nu
}=\kappa T_{\mu \nu },$ which are exactly the GR equations when $R_{\mu \nu
}=0$; i.e. GR equations and those of $f(0)=-2\Lambda $ theories are
identical for spacetimes with zero Ricci tensor.

\begin{corollary}
For spacetimes with constant Ricci scalar $R_{0},$ GR and $f(R)$ theories
such as 
\begin{equation}
R_{0}f^{\prime }\left( R_{0}\right) -2f\left( R_{0}\right) +R_{0}-4\Lambda
\neq 0,  \label{11.4}
\end{equation}%
have no common solutions.
\end{corollary}
\end{proof}

\begin{proof}
We have shown that if a metric is a solution of GR and $f(R)$ theory, the $%
f(R)$ function must satisfy the necessary and sufficient condition (\ref{006}%
), consequently it must necessarily satisfy the trace of this condition
which is given by 
\begin{equation}
R_{0}f^{\prime }\left( R_{0}\right) -2f\left( R_{0}\right) +R_{0}-4\Lambda
=0,  \label{81}
\end{equation}%
so if the $f(R)$ function doesn't satisfy this trace like in (\ref{11.4}),
the condition (\ref{006}) cannot be satisfied; so the metric cannot be
solution of the two theories.
\end{proof}

\subsection{Remarks}

1) It should be understood that the function $f(R)=R-2\Lambda $ and its
first derivative, namely the general relativity, are not the only functions
satisfying the conditions: (\ref{12}), (\ref{008}), (\ref{z}). Since $R_{0}$
is a real number, we may have an infinity of different $f(R)$ functions and
their first derivative which all have the same output $f\left( R_{0}\right) $
and/or $f^{\prime }\left( R_{0}\right) $ corresponding to the input $R_{0}$
as we will show it in what follows.

2) In the particular case where the cosmological constant is neglected, the
corollary 4 states: All the GR vacuum solutions are also vacuum solutions of
an $f(R)$ theory if and only if $f(0)=0.$ This result was also deduced by J.
D. Barrow and A. C. Ottewill\emph{\cite{bar} }using a diffrent approach with
the assumption $f^{\prime }(0)\neq 0,$ which according to our approach seems
not to be a necessary condition to prove that both theories have the same
vacuum solutions for $R_{0}=0.$

3) The conditions on the $f(R)$ function that we have obtained for which GR
and $f(R)$ theory have the same solutions when Ricci scalar is constant
should also be regarded as the conditions of existence for such theories,
indeed in the case of constant Ricci scalar we can easily see from (\ref{12}%
), (\ref{008}), (\ref{z}) that there is an infinity of functions which may
satisfy these conditions; it means that for all GR solutions such as $R$ is
constant, there exists at least one $f(R)\neq R-2\Lambda $ theory which has
the same GR solutions for the specific value of $R,$ which is not the case
if $R$ isn't constant as we will show later.

\section{The form of $f(R)$ theories equivalent to general relativity for a
specific Ricci scalar}

One may wonder what's the main interest of the previous theorem and
corollaries? In point of fact the theorem and its consequences give the
possibility to know if a given $f(R)$ theory may have the same solutions as
the general relativity ones for a specific form of the Ricci scalar. This is
very convenient if for instance we are looking for an $f(R)$ exact
solutions, where we don't necessarily need to solve $f(R)$ field equations
since we already know that the GR solutions are also solutions of the $f(R)$
theory. An added benefit is to be able to know if for a given GR solution
there exist or not $f(R)$ theories for which this GR solution is also a
solution and we may even obtain the from of the $f(R)$ functions as we are
going to see further.

\subsection{Conditions on the parameters of an $f(R)$ model}

In the literature when one talk about an $f(R)$ theory one talk specifically
about a class of functions depending on some parameters. An interesting
consequence of the theorem and its corollaries is to provide for any class
of $f(R)$ functions depending on one or many parameters the required
conditions on these parameters in such a way that the GR solutions is also a
solutions of this class in the case where $R=R_{0}$. To illustrate this
fact, let's consider the model 
\begin{equation}
f(R;\alpha ,\beta )=R+\alpha \beta ^{1-p}R^{p},\text{ with }0<p<1,\alpha
,\beta >0,  \label{29}
\end{equation}%
proposed in\cite{363} as a candidate for dark energy. From (\ref{81}), we
deduce that the parameters $\alpha $ and $\beta $ must necessarily satisfy
the condition%
\begin{equation}
\left( p-2\right) \alpha \beta ^{1-p}R_{0}^{p}-4\Lambda =0,  \label{79}
\end{equation}%
to allow the GR solution to be a solution of this model. Of course this is a
necessary condition which is insuffisant to claim that GR solutions are also
solutions of this model. On the other hand as we find that for this model,
the condition (\ref{12}) yields, 
\begin{equation}
\QATOPD\{ . {\alpha \beta ^{1-p}pR_{0}^{p-1}=0}{\alpha \beta
^{1-p}R_{0}^{p}+2\Lambda =0},  \label{80}
\end{equation}%
which means that GR and this model are equivalent for spacetimes with zero
Ricci scalar since they have exactly the same solutions; However, since $%
0<p<1;$ $\alpha ,$ $\beta >0,$ the condition (\ref{12}) is not satisfied if $%
R_{0}\neq 0$, which does not necessarily mean that the GR solution isn't a
solution for this model when $R_{0}\neq 0,$ because (\ref{12}) is just a
sufficient condition not a necessary one for having a common solutions.

From (\ref{008}), we can affirm that all vacuum solutions of Einstein's
field equations are solutions of field equations generated by $R+\alpha
\beta ^{1-p}R^{p}$ theory, if and only if 
\begin{equation}
\left( p-2\right) \alpha \beta ^{1-p}\left( 4\Lambda \right) ^{p}-4\Lambda
=0,  \label{30}
\end{equation}%
and if we neglect the cosmological constant, we deduce that all the vacuum
solutions of GR are necessarly vacuum solutions of $R+\alpha \beta
^{1-p}R^{p}$ field equations, because $\forall \alpha ,\beta ,p$, such as $%
0<p<1;$ $\alpha ,$ $\beta >0,$ we have $f(0;\alpha ,\beta )=0.$ This is the
appropriate method to be followed if we would like to know if an $f(R)$
model may or may not have the same solutions as GR.

\subsection{\protect\bigskip Construction of $f(R)$ theories equivalent to
GR when Ricci scalar is constant}

Since the required conditions making all GR solution leading to a spacetime
with constant Ricci scalar a solution of an $f(R)$ theory are known; we are
able thanks to these conditions to have more informations about the form of
such functions; in other words, starting from any GR solution it's possible
to get an infinity of $f(R)$ theories for which this solution is also an $%
f(R)$ solution.

\begin{corollary}
The vacuum solutions of GR are also the vacuum solutions of the $f(R)$
theories such as 
\begin{equation}
f\left( R\right) =\exp \left( \frac{R}{2\Lambda }\right) \left[ \frac{1}{%
\Lambda }\int F\left( R\right) \exp \left( -\frac{R}{2\Lambda }\right) dR+k%
\right] ,  \label{24}
\end{equation}%
where $k$ is an arbitrary constant and $F\left( R\right) $ is also an
arbitrary function which vanish at $R=4\Lambda .$
\end{corollary}

\begin{proof}
Let $g_{\mu \nu }$ be a vacuum solution of GR and $f$ a function of $R.$
Let's consider the linear ordinary differential equation%
\begin{equation}
\Lambda f^{\prime }\left( R\right) -\frac{1}{2}f\left( R\right) =F\left(
R\right) ,  \label{21}
\end{equation}%
where $F$ is an arbitrary function of $R$ such that $F(4\Lambda )=0;$ hence
all the solutions of this equation will satisfy the condition (\ref{008})
for $R=4\Lambda ,$ so $g_{\mu \nu }$ is a vaccum solution of the $f(R)$
theories generated by the solutions of equation (\ref{21}) i.e. the
functions of the form $f\left( R\right) =\exp \left( \frac{R}{2\Lambda }%
\right) \left[ \frac{1}{\Lambda }\int F\left( R\right) \exp \left( -\frac{R}{%
2\Lambda }\right) dR+k\right] $ because this is in fact the general solution
of (\ref{21}) where $k$ is the constant of integration.
\end{proof}

We clearly see that $f(R)=R-2\Lambda $ corresponding to the GR theory isn't
the only function satisfying the condition (\ref{008}), since this one is
just the particular case where the $F\left( R\right) $ function is equal to $%
2\Lambda -\frac{R}{2}$ amongst an infinity of others possible choices of $%
F\left( R\right) $; for instance one may choose $F\left( R\right) =0$ which
is the most simple choice; indeed this one leads to the theory $f\left(
R\right) =k\exp \left( \frac{R}{2\Lambda }\right) .$

For the general case of an arbitrary constat Ricci scalar, the task of
getting such $f(R)$ functions is a lot simpler. All we have is to suppose
any function depending on two parameters, then we impose the condition (\ref%
{12}) to the $f(R)$ function and its first derivative to obtain two
equations which will determine the exact value of the two parameters. For
instance let's consider the function 
\begin{equation}
f\left( R\right) =\alpha R^{n}+\beta R^{m},  \label{70}
\end{equation}%
where $\alpha $ and $\beta $ are two real parameters and $n,$ $m\in 
\mathbb{N}
^{\ast }(n\neq m).$ According to the condition (\ref{12}); $\alpha $ and $%
\beta $ must satisfy the system of equations 
\begin{equation}
\QATOPD\{ . {n\alpha R_{0}^{n-1}+\beta mR_{0}^{m-1}=1}{\alpha
R_{0}^{n}+\beta R_{0}^{m}=R_{0}-2\Lambda },  \label{71}
\end{equation}%
if we want to make that the GR solution $g_{\mu \nu }$ leading to $%
R_{0}=g_{^{\mu \nu }}R_{\mu \nu }$, a solution of this $f(R)$ theory. The
solution of the system of equtions (\ref{71}) is 
\begin{equation}
\QATOPD\{ . {\alpha =\frac{\left( 1-m\right) R_{0}+2m\Lambda }{\left(
n-m\right) R_{0}^{n}}}{\beta =\frac{\left( n-1\right) R_{0}-2n\Lambda }{%
\left( n-m\right) R_{0}^{m}}},  \label{72}
\end{equation}%
therefore $g_{\mu \nu }$ is also a solution of the $f(R)$ theory such as 
\begin{equation}
f\left( R\right) =\frac{\left( 1-m\right) R_{0}+2m\Lambda }{\left(
n-m\right) R_{0}^{n}}R^{n}+\frac{\left( n-1\right) R_{0}-2n\Lambda }{\left(
n-m\right) R_{0}^{m}}R^{m}.  \label{73}
\end{equation}%
There is also an option to get more classes of $f(R)$ theorie which have the
same solution of GR, by different combinations of $f^{\prime }\left(
R_{0}\right) $ and $f\left( R_{0}\right) $, enabling to obtain different
types of ordinary differential equations and in the same way these $f\left(
R\right) $ functions can be deduced by solving the obtained differential
equations. Indeed, from (\ref{12}) we can write%
\begin{equation}
f^{\prime }\left( R_{0}\right) +f\left( R_{0}\right) =R_{0}-2\Lambda +1,
\label{74}
\end{equation}%
so all the solutions of the linear ordinary differential equation 
\begin{equation}
f^{\prime }\left( R\right) +f\left( R\right) =U\left( R\right) ,  \label{75}
\end{equation}%
such that $U\left( R_{0}\right) =R_{0}-2\Lambda +1$, will necessarily
satisfy equation (\ref{74}). The general solution of (\ref{75}) is given by 
\begin{equation}
f(R)=\exp \left( -R\right) \left( \int U(R)\exp \left( R\right)
dR+c_{1}\right) ,  \label{16.1}
\end{equation}%
then we have to impose to this general solution the condition (\ref{12}), to
deduce that the constant of integration $c_{1}$ is given by 
\begin{equation}
c_{1}=\left( R_{0}-2\Lambda \right) \exp \left( R_{0}\right) -\left[ \int
U(R)\exp \left( R\right) dR\right] _{R=R_{0}}.  \label{76}
\end{equation}%
We may also take others combination of $f\left( R_{0}\right) $ and $%
f^{\prime }\left( R_{0}\right) $ in conformity with (\ref{12}), like for
instance 
\begin{equation}
f^{\prime }\left( R_{0}\right) f\left( R_{0}\right) =R_{0}-2\Lambda ,
\label{22}
\end{equation}%
then we have to solve the differential equation 
\begin{equation}
f^{\prime }\left( R\right) f\left( R\right) =Q(R),  \label{23}
\end{equation}%
where $Q(R_{0})=R_{0}-2\Lambda .$ to get the function 
\begin{equation}
f(R)=\pm \sqrt{2\int Q\left( R\right) dR+c_{2}},  \label{77}
\end{equation}%
and the condition(\ref{12}) will determine the constant of integration $c_{2}
$ which is given by 
\begin{equation}
c_{2}=\left( R_{0}-2\Lambda \right) ^{2}-2\left[ \int Q(R)dR\right]
_{R=R_{0}}.  \label{78}
\end{equation}%
This process may be repeated each time to get more $f(R)$ theories.

\subsection{Construction of the $f(R)$ functions for spacetimes with $R\neq $
Constant (case of the maximally symmetric spacetime: the FLRW solution)}

Although we have the necessary and the sufficient condition (\ref{11})
making the metric one solution for GR and $f(R)$ theory; the task of getting
the form of the $f(R)$ functions satisfying this condition for a given GR
solution isn't easy if the Ricci scalar isn't constant due to the fact that
the covariant derivative $\nabla _{\mu }\nabla _{\nu }f^{^{\prime }}\left(
R\right) $ isn't zero, so with the exception of the condition (\ref{11}) we
are not able to have others easier properties appliquable to the $f(R)$
functions as a general conditions for all the GR solutions, which may lead
to get the form of the $f(R)$ functions in a simple way like with the case
of constant Ricci scalar. Therefore this should be considered on a
case-by-case according to the concerned GR solution; keeping in mind that
contrary to the constant Ricci scalar case, there is nothing to say that
such $f(R)$ function will certainly exist for all GR solutions without any
restriction. The first step in the determination of the $f(R)$ consists
obviously to consider a GR solution such as $R$ isn't constant; for that we
will choose to investigate one of the most relevant solution of general
relativity and cosmology namely the FLRW metric by trying to answer the
question: is there any $f(R)\neq R-2\Lambda $ theory for which the FLRW is
also a solution?. The FLRW is a metric describing a maximally symmetric
spacetime. It's well known that in the spherical coordinate system, the most
general metric for the maximally symmetric spacetime is given by 
\begin{equation}
ds^{2}=-dt^{2}+a^{2}(t)\left[ \frac{dr^{2}}{1-Kr^{2}}+r^{2}\left( d\theta
^{2}+\sin ^{2}\theta d\varphi ^{2}\right) \right] ,  \label{41}
\end{equation}%
where $a(t)$ is an arbitrary function depending only on time and $K$ the
curvature of the space which take a positive value for a spherical geometry
of the space, negative for the hyperbolic case, and zero for the flat space
i. e. the euclidean space; not to be confused with Ricci scalar $R$ which is
the curvature of the spacetime. For $c=1$, the non-vanishing components of
the Ricci curvature tensor and the scalar curvature are respectively given
by 
\begin{eqnarray}
R_{00} &=&-3\frac{\ddot{a}}{a},  \label{042} \\
R_{ij} &=&\left[ \ddot{a}a+2\left( \dot{a}^{2}+K\right) \right] \sigma _{ij},
\notag
\end{eqnarray}%
\begin{equation}
R=6\left[ \frac{\ddot{a}}{a}+\left( \frac{\dot{a}}{a}\right) ^{2}+\frac{K}{%
a^{2}}\right] ,  \label{43}
\end{equation}%
where the dot denotes the derivative with respect to time; $\sigma
_{ij}dx^{i}dx^{j}=\frac{dr^{2}}{1-Kr^{2}}+r^{2}\left( d\theta ^{2}+\sin
^{2}\theta +d\varphi ^{2}\right) $; $i$ and $j$ take the values: $1,2,3$ and
if there is no confusion about the time-dependence of the function $a(t)$,
we may also use the notation $a$ as we have done. We note moreover that for
this case, when $R$ isn't constant it's only time dependant. To provide a
physical meaning to this metric, this one should satisfy Einstein field
equations and the maximally symmetric spacetime represents physically an
isotropic homogeneous universe in expansion or in contraction according to
the behavior relative to time of the $a$ function which is precisely the
scale factor of the universe. Matter, energy and radiation are described by
the stress-energy tensor $T_{\mu \nu }.$ For the perfect fluid model which
is fully consistent with an homogeneous and isotropic universe the matricial
form of the energy momentum-tensor is given by 
\begin{equation}
\left( T_{\mu \nu }\right) =\left( 
\begin{array}{cccc}
-\rho & 0 & 0 & 0 \\ 
0 & \frac{a^{2}}{1-Kr^{2}}p & 0 & 0 \\ 
0 & 0 & a^{2}r^{2}p & 0 \\ 
0 & 0 & 0 & a^{2}r^{2}\sin ^{2}\theta p%
\end{array}%
\right) .  \label{44}
\end{equation}%
$\rho $ is the energy density and $p$ the presure. Hence by imposing
Einstein field equations, we get the well-known Friedmann equations 
\begin{equation}
\left( \frac{\dot{a}}{a}\right) ^{2}+\frac{K}{a^{2}}=\frac{8\pi G}{3}\rho +%
\frac{\Lambda }{3},  \label{45}
\end{equation}%
\begin{equation}
2\frac{\ddot{a}}{a}+\left( \frac{\dot{a}}{a}\right) ^{2}+\frac{K}{a^{2}}%
=-8\pi Gp+\Lambda .  \label{46}
\end{equation}%
For all the solutions of the Friedmann equations (\ref{45}), (\ref{46}) our
objective is to find the $f(R)\neq R-2\Lambda $ theories, such as the metric(%
\ref{41}) will also be a solution of the $f(R)$ theory, if such theories
exists. In general, Friedmann equations are not exactly solvable; only few
exact solutions exist for some particular cases, thus in order to get around
the difficulty to have a general formula of the scale factor, the
appropriate approach for dealing with this difficulty is to write the
Friedmann equations in the equivalent form%
\begin{equation}
\dot{a}^{2}-\frac{\Lambda }{3}a^{2}-\frac{8\pi GN_{0}}{3}a^{-\left(
1+3\omega \right) }+K=0,  \label{47}
\end{equation}%
\begin{equation}
2a\ddot{a}+\dot{a}^{2}-\Lambda a^{2}+8\pi GN_{0}\omega a^{-\left( 1+3\omega
\right) }+K=0,  \label{48}
\end{equation}%
where we have taken into consideration the usual equation of state $p=\omega
\rho $, and the solution $\rho =\rho _{0}a_{0}^{3\left( 1+\omega \right)
}a^{-3\left( 1+\omega \right) }$ of the equation $\dot{\rho}=-3\left( \rho
+p\right) \left( \frac{\dot{a}}{a}\right) $ resulting from the energy
conservation, with $\rho _{0}:=\rho \left( 0\right) $, $a_{0}:=a\left(
0\right) $, $N_{0}=\rho _{0}a_{0}^{3\left( 1+\omega \right) }.$ Then from (%
\ref{47}), we have 
\begin{equation}
\dot{a}=\pm \sqrt{\frac{8\pi GN_{0}}{3}a^{-\left( 1+3\omega \right) }+\frac{%
\Lambda }{3}a^{2}-K}.  \label{49}
\end{equation}%
If we combine this last equation with (\ref{48}); we can write the second
derivative of the scale factor only in term of $a,$ 
\begin{equation}
\ddot{a}=-4\pi GN_{0}\left( \omega +\frac{1}{3}\right) a^{-\left( 3\omega
+1\right) }+\frac{\Lambda }{3}a.  \label{50}
\end{equation}%
In view of (\ref{49}), (\ref{50}) the Ricci scalar (\ref{43}) can also be
written only in term of the scale factor 
\begin{equation}
R=8\pi GN_{0}\left( 1-3\omega \right) a^{-3\left( \omega +1\right)
}+4\Lambda .  \label{51}
\end{equation}%
It can be seen that $\omega $ should not be equal to $-1$ and $\frac{1}{3}$,
otherwise we fall back into the case of the constant Ricci scalar that we
have already considered. If for instance $\omega =\frac{1}{3}$; the Ricci
scalar is equal to $4\Lambda ,$ it means that the metric is the maximally
symmetric vacuum solution of Einstein's field, i.e. the case of De Sitter
spacetime. The same thing for $N_{0}$ which must always be a nonzero. As $%
\omega \neq -1,$ $\frac{1}{3}$, the relation may be reversed for writing $a$
in term of $R$,%
\begin{equation}
a=\left[ \frac{R-4\Lambda }{8\pi GN_{0}\left( 1-3\omega \right) }\right] ^{%
\frac{-1}{3\left( \omega +1\right) }}.  \label{52}
\end{equation}%
The $f(R)$ functions that we are looking for must satisfy equations (\ref{11}%
). In view of the form of the metric and the Ricci scalar which is only time
dependant; the condition (\ref{11}) reads 
\begin{equation}
\left[ f^{\prime }\left( R\right) -1\right] R_{\mu \nu }-\left[ \partial
_{0}^{2}f^{\prime }\left( R\right) +\frac{1}{2}f\left( R\right) -\frac{R}{2}%
+\Lambda \right] g_{\mu \nu }-\partial _{\mu }\partial _{\nu }f^{^{\prime
}}\left( R\right) =0.  \label{53}
\end{equation}%
Due to the symmetry this condition is formed only of two independent
equations, 
\begin{equation}
\left[ f^{\prime }\left( R\right) -1\right] R_{00}+\frac{1}{2}f\left(
R\right) -\frac{R-2\Lambda }{2}=0,  \label{54}
\end{equation}%
and 
\begin{equation}
\left[ f^{\prime }\left( R\right) -1\right] \left[ \ddot{a}a+2\left( \dot{a}%
^{2}+K\right) \right] -\left[ \partial _{0}^{2}f^{\prime }\left( R\right) +%
\frac{1}{2}f\left( R\right) -\frac{R}{2}+\Lambda \right] a^{2}=0.
\label{542}
\end{equation}%
From (\ref{042}), (\ref{49}), (\ref{50}), we deduce the $00$-component of
the Ricci tensor, \ 
\begin{equation}
R_{00}=-3\left[ \frac{\Lambda }{3}-4\pi GN_{0}\left( \omega +\frac{1}{3}%
\right) a^{-3\left( \omega +1\right) }\right] ,  \label{55}
\end{equation}%
which according to (\ref{52}) may also be written in term of $R$ after
elimination of the scale factor, 
\begin{equation}
R_{00}=\frac{1}{3\omega -1}\left[ 3\left( \omega +1\right) \Lambda -\frac{%
3\omega +1}{2}R\right] ,  \label{56}
\end{equation}%
this allow to write (\ref{54}) as the first order linear differential
equation for the function $f(R)$, 
\begin{equation}
\left[ 3\left( \omega +1\right) \Lambda -\frac{3\omega +1}{2}R\right]
f^{\prime }\left( R\right) +\left( \frac{3\omega -1}{2}\right) f\left(
R\right) +\left( R-4\Lambda \right) =0.  \label{57}
\end{equation}%
From this equation we note that if the cosmological constant is negligible
and $\omega =\frac{-1}{3}$, the only solution is $f\left( R\right) =R$, i.e.
the general relativity theory. This means that there is no $f(R)$ theory
which has exactly the same scale factor as GR. On the other hand the general
solution of the equation (\ref{57}) is given by 
\begin{equation}
f(R)=\QATOPD\{ . {C\left[ \frac{3\omega +1}{2\left( 1-3\omega \right) }R+%
\frac{3\Lambda \left( \omega +1\right) }{3\omega -1}\right] ^{\frac{3\omega
-1}{3\omega +1}}+\underset{f_{RG}(R)}{\underbrace{\left( R-2\Lambda \right) }%
}\text{ if }\omega \neq -1,\frac{-1}{3},\frac{1}{3}\text{ }}{\overline{C}%
\exp \frac{R}{2\Lambda }+\underset{f_{RG}(R)}{\underbrace{\left( R-2\Lambda
\right) }}\text{ if }\omega =\frac{-1}{3},\text{ }\Lambda \neq 0},
\label{58}
\end{equation}%
where $C$ and $\overline{C}$ are two constant of integration. We confim that
general relativity theory belong to this class of theories since it's the
theory corresponding to a zero constant of integration as it must, however
these class of theories must also satisfy the second equation (\ref{542}) of
the condition (\ref{53}). As we have only two independent equations and the
first one is already satisfied; in order to satisfy the second equation it's
necessary and sufficient to satisfy the trace of the system of the two
equation i.e. the trace of the condition (\ref{53}), which is given by 
\begin{equation}
\left( \frac{\partial R}{\partial t}\right) ^{2}f^{^{\prime \prime \prime
}}\left( R\right) +\frac{\partial ^{2}R}{\partial t^{2}}f^{^{\prime \prime
}}\left( R\right) -\frac{1}{3}Rf^{^{\prime }}\left( R\right) +\frac{2}{3}%
f(R)-\frac{R-4\Lambda }{3}=0.  \label{59}
\end{equation}%
According to (\ref{49}), (\ref{50}), (\ref{51}), (\ref{52}), (\ref{58}), if $%
\omega \neq -1,\frac{-1}{3},\frac{1}{3}$, this trace reads 
\begin{equation}
C\left[ 
\begin{array}{c}
\Lambda ^{2}\left( \omega +\frac{2}{3}\right) a^{6\left( \omega +1\right)
}+4\pi GN_{0}\Lambda \left( 7\omega +\frac{17}{3}\right) a^{3\left( \omega
+1\right) } \\ 
-4\pi GN_{0}K\left( 3\omega +5\right) a^{3\omega +1}-K\Lambda \left( 3\omega
+4\right) a^{2\left( 3\omega +2\right) } \\ 
-2\left( 4\pi GN_{0}\right) ^{2}\left( 3\omega ^{2}+\omega -\frac{4}{3}%
\right)%
\end{array}%
\right] =0.  \label{60}
\end{equation}%
$C=0,$ is the obvious results since it corresponds to the GR case (GR
solutions is certainly a GR solution); it's the term between brackets which
should vanish if it exists at least one $f(R)\neq R-2\Lambda $ theory for
which the GR solution is also a solution. As the scale factor is not
constant this term can not be identically zero for all $\omega ,$ $K,$ $%
\Lambda $. Yet one can note that if the cosmoligical constant is zero and
the space is flat ($\Lambda =K=0$), the trace (\ref{60}) vanishes, if 
\begin{equation}
3\omega ^{2}+\omega -\frac{4}{3}=0,  \label{61}
\end{equation}%
namely if 
\begin{equation}
\omega =\frac{-1\pm \sqrt{17}}{6}.  \label{62}
\end{equation}%
For the second case where $\omega =\frac{-1}{3},$ $\Lambda \neq 0$, the
trace (\ref{59}), is given by 
\begin{equation}
\overline{C}\left[ \frac{1}{3}a^{4}+\frac{1}{\Lambda }\left( \frac{40\pi
GN_{0}}{3}-3K\right) a^{2}+\frac{16\pi GN_{0}}{\Lambda ^{2}}\left( \frac{%
8\pi GN_{0}}{3}-K\right) \right] =0,  \label{63}
\end{equation}%
which means that the theory $f\left( R\right) $ such as $f\left( R\right) =$ 
$\overline{C}\exp \frac{R}{2\Lambda }+\left( R-2\Lambda \right) $ and $%
\overline{C}\neq 0$ cannot have exactly the same scale factor as that of GR.
It's worth to point out that these results are completely independent from
the sign of for the first derivative of the scale factor in (\ref{49}) which
has been taken positive, because if we redo the same calculation with the
negative value, we will lead to the same conclusion: For the homogeneous and
isotropic universe the only $f(R)$ theory which has exactly the same scale
factor as GR is that for 
\begin{equation}
f(R)=C\left[ \frac{1}{2}\left( \frac{1\pm \sqrt{17}}{3\mp \sqrt{17}}\right) R%
\right] ^{\frac{\pm \sqrt{17}-3}{\pm \sqrt{17}+1}}+R,  \label{83}
\end{equation}%
if the space is flat with no cosmological constant and an equation of state: 
\begin{equation}
p=\frac{-1\pm \sqrt{17}}{6}\rho ,  \label{84}
\end{equation}%
otherwise the two theories cannot have exactly the same scale factor.

\section{\textbf{Conclusion }}

Finding an equivalence between general relativity and $f(R)$\ modified
theory of gravity in the meaning that the two theories have exactly the same
mathematical solutions for spacetimes with specific scalar curvature,
doesn't really need a complicated mathematical mechanism if the Ricci scalar
is constant; indeed as we have shown, the main idea in order to achieve this
aim is to bring back Einstein's field equations inside $f(R)$\ ones(\ref{40}%
) which enable to obtain the necessary and the sufficient condition on the
function $f(R)$\ (\ref{006}) in a way that the equivalence becomes an
evidence. The consequence is to get some conditions on the function $f(R)$\
and its first derivative(\ref{12}), (\ref{008}), (\ref{z}), allowing by
different manners the obtainment of the $f(R)$ equivalent theories(\ref{24}%
),(\ref{73}), (\ref{16.1}), (\ref{77}), giving also the possibility for
fixing the conditions on the parameters of any $f(R)$\ model so that this
one will be equivalent to GR, when Ricci scalar is equal to a certain
constant as we have seen in (\ref{79}), (\ref{30}) for the dark energy model
(\ref{29}) where getting an exact solutions is a hard task even for the
vacuum case, whereas with the help of the criterion (\ref{008}), without
being obliged to solve the complicated $f(R)$ field equations, we are able
to predict that GR vacuum solutions are also a vacuum solution for this
model. In general relativity, astrophysics and cosmology, dealing with
solutions leading to spacetimes with constant Ricci scalar is a relevent
issue, at least because the best description of black holes in the universe
is provided by means of Schwarzschild and De Sitter solutions, furthermore
the interest for the common solutions GR-modified theory can be directly
highlighted for concrete physical phenomenons like the collapse of stars in
the formation of a black holes as proven by Stephen Hawking for the theory
of Brans Dicke\cite{ha} when he showed that a stationary space containing
black hole is solution of the Brans-Dicke field equations if and only if
it's a solution of Einstein field equations. Although the generalisation of
the issue is complicated we didn't restricted the invistigation only to the
case of constant scalar Ricci scalar; since we have also considered the
relevant case of the FLRW solution involving a time dependent Ricci scalar,
when we have shown that the only class of the $f(R)$ theories which has the
same scale factor as GR, is of the form (\ref{83}) valid only for the case
of a flat space with no cosmological constant and the equation of state (\ref%
{84}). This important result come in addition to many works\cite{a}\cite{b}%
\cite{c}\cite{d} dealing with FLRW metric in $f(R)$ theory performed during
these last years.

Whatever the considerations; understanding the relation between GR and any
modified theory of gravity is the only way to know if the modification has
reached the objective for which it has been established i.e. that of going
beyond general relativity.

\end{document}